# Janus Polymeric Giant Vesicles on Demand: A Predicting Phase Separation Approach for Efficient Formation

Eloise Equy,[a,b] Emmanuel Ibarboure,[a] Eric Grelet,[*,b] Sébastien Lecommandoux[*,a]

AUTHOR ADDRESS

[a] Univ. Bordeaux, CNRS, Bordeaux INP, LCPO, UMR 5629, F-33600 Pessac, France.
[b] Univ. Bordeaux, CNRS, CRPP, UMR 5031, F-33600 Pessac, France.

**ABSTRACT:** Janus particles, with their intrinsic asymmetry, are attracting major interest in various applications, including emulsion stabilization, micro/nano-motors, imaging and drug delivery. In this context, Janus polymersomes are particularly attractive for synthetic cell development and drug delivery systems. While they can be achieved by inducing a phase separation within their membrane, their fabrication method remains largely empirical. Here we propose a rational approach, using Flory-Huggins theory, to predict self-assembly of amphiphilic block copolymers into asymmetric Janus polymersomes. Our predictions are experimentally validated by forming highly stable Janus giant unilamellar vesicles (JGUVs) with a remarkable yield exceeding 90% obtained from electroformation of various biocompatible block copolymers. We also present a general phase diagram correlating mixing energy with polymersomes morphology, offering a valuable tool for JGUVs design. These polymersomes can be extruded to achieve quasi-monodisperse vesicles while maintaining their Janus-like morphology, paving the way for their asymmetric functionalization and their use as active carriers.

## INTRODUCTION

Janus particles, named after the two-faced Roman god Janus, are characterized by two distinct sides often composed of different materials and exhibiting different properties or functionalization. They are engineered into various shapes (sphere, rod, …) using inorganic (metal, silica, …) or organic (polymer, lipid, …) materials.[1–9] These asymmetric structures have attracted considerable interest across multiple fields. They can serve as stabilizers for Pickering emulsions,[1,4,6] are utilized in the development of electrophoretic inks[4,6] and contribute to the fabrication of modulated optical nanoprobes (MOONs).[1,4,10] Additionally, they can self-assemble into superstructures enabling the bottom-up design of new functional materials.[4,6] Symmetry breaking is also an essential aspect for inducing self-propulsion in a system. Janus morphology is used in the development of micro/nano-motors capable of navigating autonomously by creating a local gradient or field.[2,4,6] These systems exhibit active motion often driven by chemical-based diffusiophoresis or by external stimuli including light, electric and magnetic fields.[11–24] Micro/nano-swimmers present opportunities for applications ranging from pollutant removal[25,26] to enhancing drug delivery efficiency to desired sites within the body.[12,27,28] Moreover, Janus particles exhibit favorable properties for biomedical applications combining, bio-imaging, sensing, drug delivery and active targeting.[1,2,5] They enable dual-imaging,[29] loading of multiple drugs[30,31] and drugs in combination with contrast agents.[32,33] Moreover, Janus particles presenting distinct domains on their surface show different interactions with their environment or biological entities compared to homogeneous, symmetric particles.[5,34–36] Other studies also explore the potential of Janus particles for tissue engineering and wound sealing.[1,37]

Among the many Janus particles developed, Janus vesicles, resulting from the self-assembly of lipids or amphiphilic polymers in bilayer membranes with two distinct sides, are particularly interesting in the context of drug delivery.[38–42] Indeed, both hydrophilic and hydrophobic drugs can be loaded respectively in the lumen or in the membrane of the vesicle.[39,41,40,42,38] Smart multi-component carriers can be designed using Janus vesicles. One could imagine selectively loading therapeutic agents on each side based on their affinity with various polymers and releasing them at different times by tuning polymer permeability or in a controlled manner using polymer responsive to various stimuli, such as pH, temperature, etc.[43–52] Inorganic nanoparticles can also be incorporated as contrast agent for imaging or for photothermal therapy [53–61]. Similarly to Janus particles, vesicles can also self-propel.[62–65] Furthermore, vesicles showing lateral phase separation offer considerable potential for fabricating synthetic cells, which in addition to explore novel applications, can serve as a tool to study some fundamental processes of natural cells.[66–72] Indeed, they mimic the inhomogeneous cell membrane containing lipid raft domains associated with important biological processes including endocytosis, adhesion, signaling, protein transport and apoptosis.[73,74] Synthetic vesicles provide a platform for investigating protein-driven phase separation and protein sequestration within domains.[75] Moreover, vesicles domains can serve as molecular channel transporting small molecules across the membrane.[76] Additionally, Janus vesicles can promote vesicles clustering,[77] membrane fusion[78–82] and fission.[83–90]

Since the initial discovery of polymersomes,[91,92] various examples of Janus polymersomes are described in the literature and prepared using different strategies.[53,60,62,86,88,89,93–115] One common approach is based on the induction of a phase separation within vesicle membrane between incompatible amphiphiles. The formation of Janus liposome has already been widely studied over the last decades.[116–120] Lipidic phase separation into liquid-ordered and liquid-disordered domains can be obtained by mixing a high melting temperature (saturated) lipid with a low melting temperature (usually unsaturated) lipid, and a sterol.[64,77,119,121] With the aim of modulating membrane properties, researchers have also tried to achieve phase separation in lipid/polymer hybrid vesicles.[122] Janus Giant Hybrid Unilamellar Vesicles (JGHUVs) as well as patchy or stripped vesicles have been successfully prepared by mixing different lipids and various copolymers with or without cholesterol.[88,89,93,95,96,123,124] However, most of lipid/polymer JGHUVs suffer from lack of stability and tend to undergo fission into distinct liposomes and polymersomes, especially when mixed with fluid lipids.[88,89,96] The main reason stems from the strong incompatibility and hydrophobic mismatch between polymers and lipids, leading to asymmetry due to kinetically trapped and non-thermodynamically stable structures. Polymer/polymer phase separation within vesicles could be an alternative to obtain more stable systems.[125,126] Battaglia and coll. have widely studied polymersomes with inhomogeneous membrane at the submicrometer scale,[36,62,97,127–129] emphasizing how variations in hydrophobic block compositions appear to be the main driving force for polymer segregation.[129] Using two incompatible block copolymers they successfully formed phase separated budded large unilamellar vesicles (LUV, diameter range: 100-1000 nm).[62] Similarly, Landfester and coll. prepared Janus giant unilamellar vesicles (JGUVs) and demonstrated the need to use a polymer composed of the two hydrophobic blocks as compatibilizer to prevent fission.[98] The importance of adding a third compatibilizing copolymer to stabilize JGUVs was also supported by simulation studies.[83,130]

More recently the effect of membrane phase separation due to a shape transformation was studied using lower critical solution temperature (LCST) polymers.[86] However, while most investigations focus on phase separation driven by differences in the hydrophobic block, phase separation can also be achieved between two copolymers with the same (or different) hydrophobic block by using different hydrophilic segments, particularly using charged hydrophilic blocks.[100,131] Altogether, despite the strong interest in asymmetric Janus like vesicles, the systems and methods allowing their preparation are rather empirical and often lead to systems that are unstable or produced in low yield.

In this study, we present a generic model based on Flory-Huggins theory to guide and rationalize the selection of copolymers suitable for forming Janus giant unilamellar vesicles (JGUVs) by polymer phase separation. Our investigation focuses on understanding the key thermodynamic parameters driving phase separation. The validity of our model is experimentally confirmed by probing the effects of various block copolymer compositions and sizes, along with temperature. The efficiency of our approach results in substantial amounts of highly stable, nearly monodisperse JGUVs. This enables us to propose a comprehensive phase diagram correlating mixing energy with polymersome morphology, thereby opening the way for the design of a wide variety of Janus polymersomes.

## RESULTS & DISCUSSION

To develop a predictive model for Janus giant unilamellar vesicle (JGUVs) formation, we hypothesize that the primary driving force is the phase separation between the hydrophobic blocks forming the membrane, as the hydrophilic PEG block is common in all investigated copolymers.[98,129] We chose PEG as a hydrophilic segment because it is broadly used in many different applications. Based on this assumption, we explore the applicability of the Flory-Huggins theory, a well-established thermodynamic framework for describing polymer phase separation in bulk. In this context, we aim to apply this theory to rationally select polymers based on their phase separation behavior.

### Theoretical prediction of polymer miscibility through calculation of Gibbs free energy of mixing

In the Flory-Huggins model, polymer-polymer miscibility is driven by the free energy of mixing ($\Delta G_m$) which comprises two antagonist contributions: an enthalpic component ($\Delta H_M$) unfavourable to the mixing and an entropic component ($\Delta S_M$) promoting the mixing:

$$\Delta G_m = \Delta H_M - T\Delta S_M \quad \text{Eq. 1}$$

Here, $\Delta G_m < 0$ indicates miscible polymers, while $\Delta G_m > 0$ implies immiscible polymers.

For a mixture of polymers A and B, the molar Gibbs Free energy of mixing is expressed as follow:

$$\frac{\Delta G_m}{n} = RT\left(\frac{\Phi_A}{DP_A}ln\Phi_A + \frac{\Phi_B}{DP_B}ln\Phi_B + \chi_{AB}\Phi_A\Phi_B\right) \quad \text{Eq. 2}$$

with $n$ the total number of moles of monomer units, $R = N_A k_B = 8.314$ J.mol$^{-1}$.K$^{-1}$, $N_A$ being Avogadro's number and $k_B$ Boltzmann constant, $\Phi_A$ and $\Phi_B$ the volume fractions, $DP_A$ and $DP_B$ their respective degree of polymerization and $\chi_{AB}$ the Flory interaction parameter, which is a dimensionless quantity quantifying the interaction strength between different polymer segments:

$$\chi_{AB} = \frac{v_{AB}}{RT}(\delta_A - \delta_B)^2 \quad \text{Eq. 3}$$

with $v_{AB} = \sqrt{V_m^A V_m^B}$ the geometric mean of polymers A and B molar volumes, in cm$^3$.mol$^{-1}$, and $\delta$ their respective solubility parameters in (J/cm$^3$)$^{1/2}$.

The Hildebrand solubility parameters have been calculated using the following equation: [132]

$$\delta = \sqrt{\frac{E_{coh}}{V}} \quad \text{Eq. 4}$$

The cohesive energy and the molar volume have been obtained using Fedors group contribution method for its simplicity and broad applicability, including silicon as chemical moiety.[133] Data are reported in Supporting Information: Tables S1, S2, S3, S4. The calculated solubility parameter values for each polymer used are reported in Table 1.

**Table 1. Block copolymer characteristics.**

| Polymer | $Mn_{hydrophobic}^{a}$ (g/mol) | $Mn_{hydrophilic}^{a}$ (g/mol) | $f_h^{b}$ | $Đ^{c}$ | $δ^{d}$ $(J/cm^3)^{1/2}$ |
|---|---|---|---|---|---|
| PBD$_{22}$-b-PEG$_{14}$ | 1200 | 600 | 0.33 | 1.09 | 17.25 |
| PTMC$_{51}$-b-PEG$_{22}$ | 5200 | 1000 | 0.16 | 1.03 | 22.72 |
| PDMS$_{23}$-b-PEG$_{13}$ | 1700 | 600 | 0.26 | 1.12 | 15.11 |
| PDMS$_{27}$-b-PEG$_{17}$ | 2000 | 750 | 0.26 | 1.11 | 15.11 |
| PDMS$_{36}$-b-PEG$_{23}$ | 2700 | 1000 | 0.27 | 1.05 | 15.11 |

[a] Number average molecular weight (Mn) and degree of polymerization (DP) determined by NMR; [b] Hydrophilic fraction defined as Mn(hydrophilic block)/[Mn(hydrophilic block)+ Mn(hydrophobic block)]; [c] Dispersity determined by chromatography (SEC); [d] Solubility parameter calculated using Fedors group contribution method reported in Supporting Information.

In a nutshell, $\Delta G_m/n$ as expressed in Eq.2, depends on four key parameters:

- Polymer chemistry: The chemical groups on the polymer backbone define the solubility parameters, which in turn determine the Flory interaction parameter, χ.
- Composition of the system through the relative volume fraction $\Phi$ of each component.
- Polymer chain length *via* the degree of polymerization.
- Temperature, as it directly appears in Eqs. 2 and 3.

We investigate how these parameters influence polymer miscibility and assess whether the Flory-Huggins theory is effective in predicting phase separation within polymersome membranes. To do so, we selected several polymers able of forming polymersomes, either commercially available or synthesized in the laboratory, all possessing a common hydrophilic PEG segment (Table 1, Scheme 1).[93,134,135] From these polymers, we calculated the molar Gibbs free energy of mixing (ΔG$_m$/n) at different composition between their hydrophobic blocks pairwise, using Eq.2.

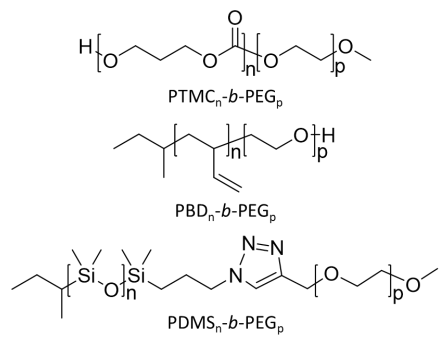

**Scheme 1. Structure of the copolymers used in this study.**

To verify our hypothesis, giant unilamellar vesicles (GUVs, diameter range: 1–200 μm) are concurrently prepared by electroformation (Figure 1). Briefly, solutions in chloroform are prepared by mixing two copolymers at various ratios. These solutions are spread onto conductive glass slides and solvent is evaporated to form a dry polymer film. The film is then rehydrated with a sucrose solution while applying a sinusoidal electric field. After 1 hour, vesicles are collected. To investigate their morphology by confocal microscopy and differentiate the polymers in the membrane, labelled polymers with fluorescent dyes are initially incorporated in the chloroform solutions.

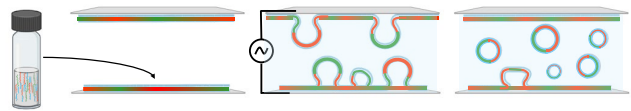

Figure 1. Schematic representation of the preparation of Janus polymersomes by electroformation (red and green parts represent schematically different polymer chains).

### Influence of polymer nature and composition on polymersome morphology

Analysis of the phase diagram in Figure 2A and D allows us to anticipate that mixing PBD$_{22}$-b-PEG$_{14}$ with either PTMC$_{51}$-b-PEG$_{22}$ or PDMS$_{27}$-b-PEG$_{17}$ suggests favorable conditions for inducing phase separation within the membrane. Indeed, they exhibit a positive value of the Gibbs free energy, $\Delta G_m$, that is characteristic of an incompatible system, but with different order of magnitude. Vesicles are prepared using polymer solutions at 55 and 23 v% of PBD block for a mixture of PBD$_{22}$-b-PEG$_{14}$ with PTMC$_{51}$-b-PEG$_{22}$, as well as, 50 and 80 v% of PBD for a mixture of PBD$_{22}$-b-PEG$_{14}$ with PDMS$_{27}$-b-PEG$_{17}$. Fluorescence microscopy observations of the polymersomes reveal that PTMC$_{51}$-b-PEG$_{22}$ and PBD$_{22}$-b-PEG$_{14}$ form distinct vesicles either composed solely of PTMC-b-PEG in pink or pure PBD-b-PEG in red, regardless of the ratio (Figure 2B, 2C and Figure S2). Consequently, one can deduce that these polymers are too incompatible and form vesicles independently of each other. In contrast, the mixture of PBD$_{22}$-b-PEG$_{14}$ and PDMS$_{27}$-b-PEG$_{17}$ evidences a theoretical phase diagram with much smaller values of $\Delta G_m/n$ (about 100 times smaller), and even negative values at some compositions, hinting at possible compatibility. This suggests that a 50% volume fraction of PBD might be an ideal scenario, with a low degree of incompatibility ($\Delta G_m/n > 0$, but small value) that could promote phase separation within the membrane. This is experimentally confirmed by the visual observation of red domains (PBD) and green domains (PDMS) within the vesicle membranes (Figure 2E and Figure S3). At 50 v% of PBD, polymers exhibit enough incompatibility to phase separate, but not to the extent of forming entirely independent vesicles, as observed in the PBD / PTMC system. Moreover, since the chemical incompatibility is low, no compatibilizer is needed to prevent complete fission, as used by Rideau and coll.[98] or in computational design of JGUVs.[83,130] Conversely, at 80 v% of PBD (*i.e.*, $\Delta G_m/n < 0$), the membrane exhibits a homogeneous mixture of the two copolymers, with overlaid green and red signals (Figure 2F and Figure S4). These observations are fully consistent with a mixture of two compatible blocks, in quantitative agreement with our prediction. Previous studies on different systems already evidenced that different morphologies of phase separated vesicles can be obtained by varying composition.[93,116,129]

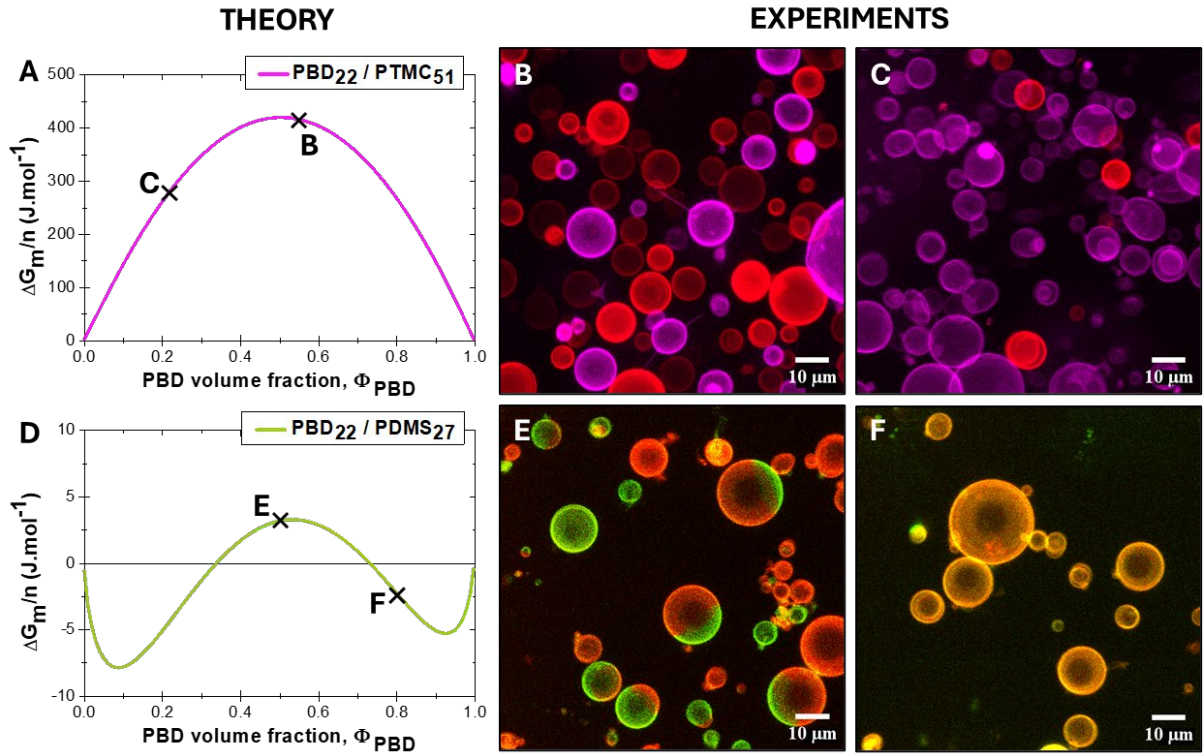

Figure 2. Influence of the chemical structure and composition on the GUV phase separation. $\Delta G_m/n$ dependence with the volume fraction of PBD for PBD$_{22}$-b-PEG$_{14}$ mixed with (A) PTMC$_{51}$-b-PEG$_{22}$ or (D) with PDMS$_{27}$-b-PEG$_{17}$. CLSM 3D reconstruction images of polymersomes prepared by electroformation. (B) 55 v% of PBD hydrophobic bloc ≡ 50 w% of PBD$_{22}$-b-PEG$_{14}$ (labelled in red) mixed with PTMC$_{51}$-b-PEG$_{22}$ (labelled in pink), forming completely separated polymersomes. (C) 23 v% of PBD hydrophobic bloc ≡ 20 w% of PBD$_{22}$-b-PEG$_{14}$ (labelled in red) mixed with PTMC$_{51}$-b-PEG$_{22}$ (labelled in pink), forming completely separated polymersomes. (E) 50 v% of PBD hydrophobic bloc ≡ 50 w% of PBD$_{22}$-b-PEG$_{14}$ (labelled in red) mixed with PDMS$_{27}$-b-PEG$_{17}$ (labelled in green), forming Janus polymersomes. (F) 80 v% of PBD hydrophobic bloc ≡ 80 w% of PBD$_{22}$-b-PEG$_{14}$ (labelled in red) mixed with PDMS$_{27}$-b-PEG$_{17}$ (labelled in green), forming homogeneously mixed polymersomes. Scale bar: 10 μm.

### Influence of the degree of polymerization on Janus polymersome formation

Similarly, the impact of degree of polymerization (DP) of the hydrophobic polymers, influencing the entropy of mixing, is predicted by calculating the molar Gibbs free energy of the mixtures. We use the same PBD$_{22}$-b-PEG$_{14}$ in combination with PEG-b-PDMS with different DP (Figure 3). Compared to the previous system using PDMS$_{27}$-b-PEG$_{17}$ (DP 27), lower molar mass of PDMS (DP 23) results in negative values of $\Delta G_m/n$, indicating compatibility between the polymers suggesting the formation of homogeneous mixed vesicles whatever their composition. In contrast, a higher molar mass of PDMS (DP 36) leads to a more positive $\Delta G_m/n$, meaning an increased polymer incompatibility and potential vesicle fission. Experimental results align perfectly with these predictions, as demonstrated in Figure 3 for mixtures containing PDMS with these three different DPs at 50 v% of PBD block. On the one hand, PBD$_{22}$-b-PEG$_{14}$ with the smaller PDMS$_{23}$-b-PEG$_{13}$ self-assembles into homogeneous vesicles composed of both polymers, visually represented in yellow in confocal microscopy images (Figure 3B, Figure S5). On the other hand, the larger PDMS$_{36}$-b-PEG$_{23}$ primarily produces unmixed vesicles, observed as distinct green and red entities in microscopy images (Figure 3D, Figure S6). Remarkably, only systems with $\Delta G_m/n$ >0 but relatively small, i.e. in the range 3 J.mol$^{-1}$ close to room temperature, seem to efficiently form Janus vesicles (Figure 3C). These observations also highlight how the control of the polymerization process and polymer chain dispersity are critical and need to be accurately controlled.

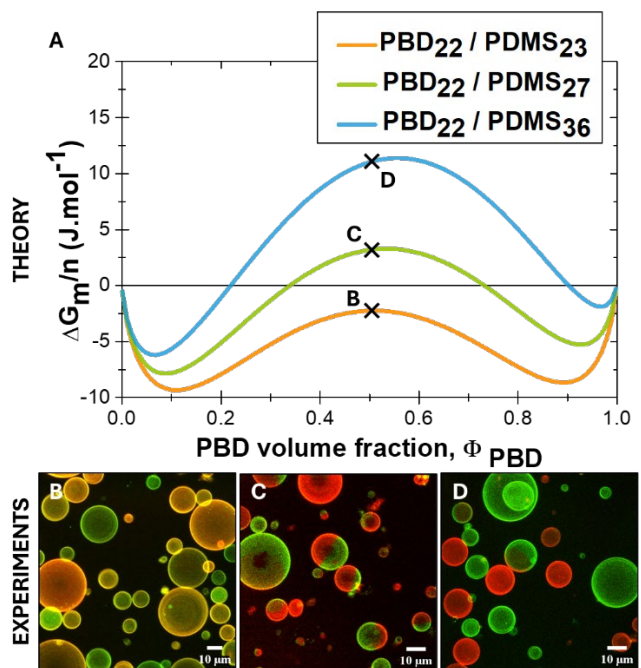

**Figure 3. Dependence of the degree of polymerization on the GUV phase separation.** (A) $\Delta G_m/n$ as function of the volume fraction of PBD for mixtures of PBD$_{22}$-*b*-PEG$_{14}$ and different sizes of PDMS-*b*-PEG: (B) PDMS$_{23}$-*b*-PEG$_{13}$, (C) PDMS$_{27}$-*b*-PEG$_{17}$, (D) PDMS$_{36}$-*b*-PEG$_{23}$. CLSM 3D reconstruction images of polymersomes prepared by electroformation from the corresponding mixtures of copolymers at 50 v% of PBD. PBD-*b*-PEG is labelled in red and PDMS-*b*-PEG is labelled in green. Scale bar: 10 μm.

### Influence of temperature on polymersome morphology

The influence of temperature on phase separation within the polymersome membrane has been extensively studied for systems containing high melting temperature lipids,[95,123] but less attention has been paid to its effect on fully polymeric vesicles. Given that the Flory interaction parameter $\chi$ scales inversely with temperature (Eq. 3), a decrease in temperature is expected to theoretically increase incompatibility, and *vice versa*. This phenomenon becomes clear when calculating $\chi$ and $\Delta G_m/n$ at different temperatures for PBD$_{22}$-*b*-PEG$_{14}$ and PDMS$_{27}$-*b*-PEG$_{17}$ mixture (Figure 4A). Indeed, $\Delta G_m/n$ is negative at 60°C, indicating compatibility. However, at 4°C, $\Delta G_m/n$ becomes positive and surpasses the curve at 20°C previously described (Figure 2D), suggesting enhanced incompatibility. Vesicles are prepared with a mixture of PBD$_{22}$-*b*-PEG$_{14}$ and PDMS$_{27}$-*b*-PEG$_{17}$ at 50 v% of PBD block using electroformation at 60°C. Fluorescence microscopy reveals the presence of patchy vesicles immediately after formation at this high temperature (Figure 4B1). These patches represent domains of each polymer within the initial vesicle membrane showing that polymers are more compatible and tend to mix. By cooling down these vesicles to 4°C for 24 hours, the incompatibility between the polymer blocks increases, leading to patch coalescence to limit interfaces, and the formation of vesicles with two single domains (Figure 4C1). This effect is reversible, as demonstrated by the possibility of obtaining patchy vesicles again by re-heating the JGUVs (Figure 4B2), which resume their Janus morphology after a few hours at low temperature (Figure 4C2). This approach mitigates the risk of forming vesicles composed solely of one type of polymer due to pronounced phase separation in the film or differences in the kinetics of formation between copolymers, likely influenced by their glass transition temperatures (Tg). Using this method, we have successfully increased the yield of Janus polymersomes production (Figure S7 & Video S1 and S2). When PBD$_{22}$-*b*-PEG$_{14}$ and PDMS$_{27}$-*b*-PEG$_{17}$ are used at 50 v% of PBD, 93 ± 3 % of polymersomes formed exhibit Janus-like morphology when electroformed at 60°C (N=2846, over five different experiments), compared to only 56 ± 10 % (N=218 over two different experiments) when vesicles are prepared at room temperature. These JGUVs are quite stable over time, as their morphology remains unchanged even after 6 months of storage (Figure S9). Notably, they do not undergo fission over time, a phenomenon commonly observed in hybrid lipids/polymer vesicles.[88]

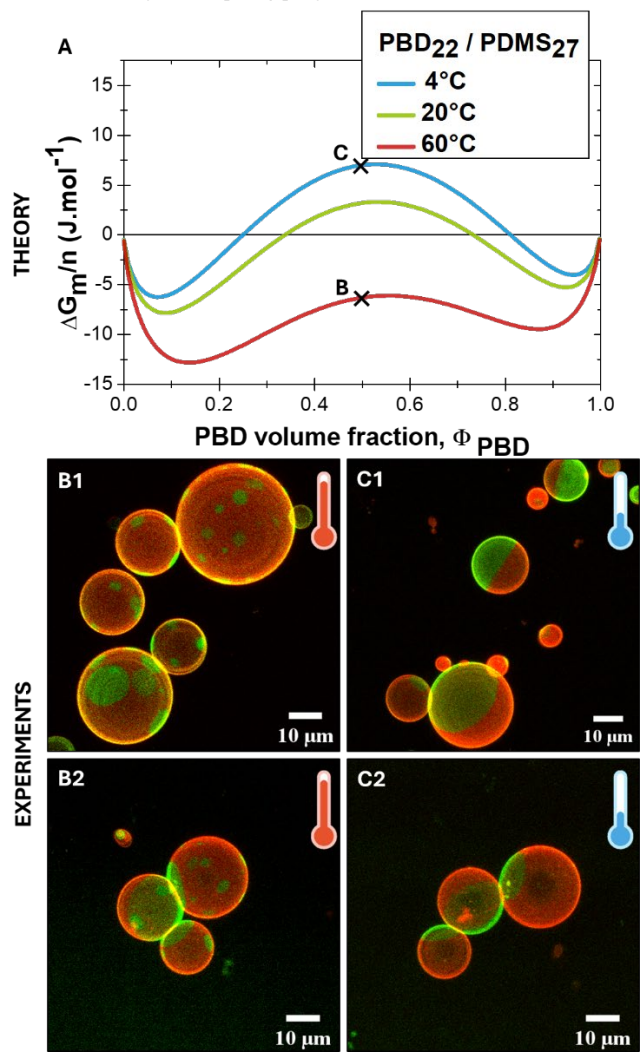

**Figure 4. Role of temperature on the GUV phase separation.** (A) $\Delta G_m/n$ dependence on the volume fraction of PBD for PBD$_{22}$-*b*-PEG$_{14}$ mixed with PDMS$_{27}$-*b*-PEG$_{17}$ at different temperatures. CLSM 3D reconstruction images of polymersomes prepared by electroformation with 50 v% of PBD hydrophobic bloc ≡ 50 w% of PBD$_{22}$-*b*-PEG$_{14}$ copolymer mixed with PDMS$_{27}$-*b*-PEG$_{17}$ at 60°C: (B1) image taken just after electroformation showing patchy vesicles (C1) image taken after 24 hours at 4°C, showing Janus vesicles. (B2) image of JGUVs re-heated 1h at 60°C, showing patchy vesicles. (C2) image of vesicles re-cooled, showing Janus vesicles. PBD-b-PEG is labelled in red and PDMS-b-PEG is labelled in green. Scale bar: 10 μm.

## Toward monodisperse vesicles

Electroformation proves to be an efficient method for producing GUVs; however, they tend to exhibit considerable polydispersity in size. To reduce the polydispersity of our Janus polymersomes, we leverage the influence of temperature on the membrane morphology and use a post-formulation extrusion process to resize vesicles to a desired uniform size. The extrusion method has previously been used to produce Janus liposomes with a diameter of 5 μm.[64] However, we anticipate that our approach, which starts with patchy vesicles, will achieve a higher yield of vesicles exhibiting phase-separated domain. Therefore, vesicles prepared at 60°C, resulting in more uniform patchy membranes, are subsequently extruded through a 5 μm membrane at the same elevated temperature to produce highly monodispersed JGUVs of ~ 5 μm in diameter. As shown in Figure 5C, both the size and polydispersity of the sample decrease after extrusion: the mean vesicle diameter after extrusion is reduces to 5 ± 2 μm (N=555, over two different experiments). It is worth mentioning that the Janus morphology is successfully preserved during the high temperature extrusion process, as shown in Figure 5A, B and Figure S8. We measure that 94 ± 3 % of the polymersomes are Janus-like before extrusion (N= 941, over two different experiments) and this percentage is identical after extrusion with 95 ± 3 % (N= 970, over two different experiments). Extruded JGUVs also demonstrate remarkable stability over at least two months (Figure S9). This extended stability makes them suitable for various applications or functionalization, even weeks after their formation. However, it should be noted that only vesicles sufficiently large enough to sediment at the bottom of the observation chamber due to sucrose / glucose density difference, and subsequently be visualized by optical microscopy (limited by the optical resolution), have been considered. Smaller polymersomes are likely present in the sample, and their morphology could differ from the observed GUVs. While the present conditions are relevant for biological purposes and osmotic pressure regulation is known to be critical for vesicle stability, further studies should be conducted to assess the formation of sub-micrometer sized vesicles by hydration, as well as their corresponding morphology.

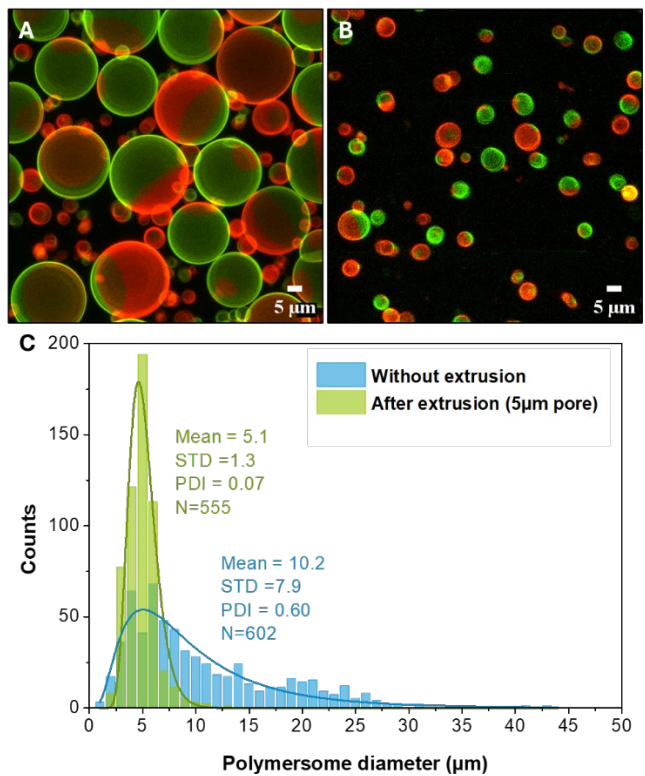

**Figure 5. Decreasing JGUV size and polydispersity.** CLSM 3D reconstruction images of polymersomes prepared by electroformation with 50 w% mixture of $PBD_{22}$-b-$PEG_{14}$ and $PDMS_{27}$-b-$PEG_{17}$ at 60°C (50 v% of PBD hydrophobic bloc). (A) Pristine vesicles after one night at 4°C. (B) Vesicles extruded through a 5 μm pore size membrane at 60°C. PBD-b-PEG is labelled in red and PDMS-b-PEG is labelled in green. Scale bar: 5 μm. (C) Size distributions of vesicles before and after extrusion through a 5 μm membrane and their corresponding lognormal fit. N represent the number of vesicles measured. Mean value, standard deviation (STD) and polydispersity index (PDI) are calculated using the equations provided in the Supporting Information.

## Predictive phase diagram

By collecting all our experimental data, we have constructed a phase diagram combining the calculated Gibbs free energy values and the obtained morphologies, which is available in the Supporting Information as Figure S10. By analyzing these data, we are able to generalize our observations and propose a generic phase diagram that allows for quantitative predictions of the vesicle morphology obtained when forming vesicles by electroformation method upon mixing of block copolymers, able to form vesicles on their own, as depicted in Figure 6. To form JGUVs by hydration method, two hydrophobic block copolymers should be mixed at a composition of 30 to 70 v% of one hydrophobic block with a corresponding Gibbs free energy of mixing between 1 and 5 J.mol$^{-1}$. For mixtures associated with a negative value of $\Delta G_m$ at room temperature, both polymers homogeneously mix within polymersome membrane. Oppositely, for mixtures leading to $\Delta G_m/n > 10$ J.mol$^{-1}$, complete phase separation occurs, vesicles composed solely of one type of polymer are obtained. In between these regimes, different morphologies coexist. As foreseen by Fraaije *et al.*[130] the range of energy that enables the formation of JGUVs, where the incompatibility between the hydrophobic blocks is strong enough for phase separation but not yet for splitting the assembly, is relatively narrow when the hydration method is used.

This is likely due to the fact that phase separation might already occurs in the polymer film during drying process,[98] which restricts the likelihood of forming vesicles that contain both polymers. The formation of JGUVs can be improved by employing strategies such as formulation at high T, which favors the mixing of polymers in the film and during the hydration process.

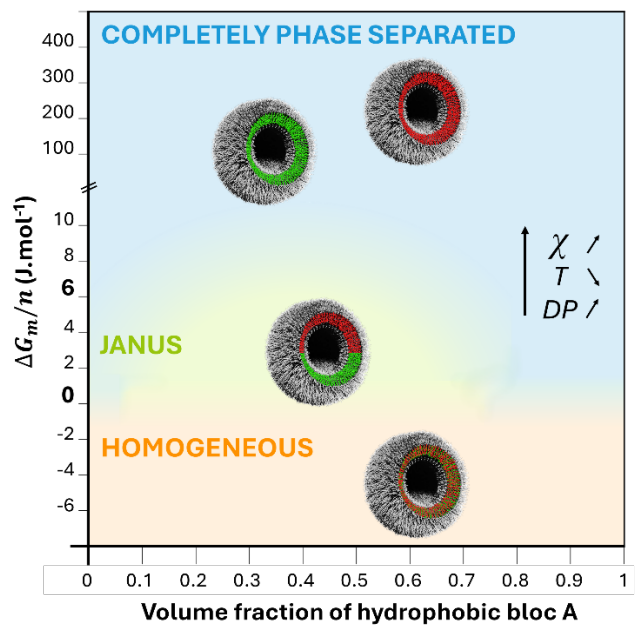

**Figure 6. General phase diagram** for the formation of JGUVs indicating the typical Gibbs free energy of mixing associated with the different polymersome morphology.

Furthermore, we can calculate the molar Gibbs free energy of the system studied by Rideau *et al*.[98] composed of PBD$_{73}$-*b*-PEEP$_{12}$ and PDMS$_{61}$-*b*-PEEP$_{12}$ to determine if our approach can account for the morphology they observed through film hydration. The molar Gibbs free energy associated with their system at 30, 50 and 70% of PBD-*b*-PEEP are 39 J.mol$^{-1}$, 47 J.mol$^{-1}$ and 39 J.mol$^{-1}$ respectively. These values align with their observations of fully separated vesicles composed exclusively of either PBD or PDMS in the absence of a compatibilizer.

### Potential limitations

It is important to mention that this approach has some limitations, primarily because it neglects potential entropic constraints of polymer chains when confined within the vesicle membrane. However, since we are mostly working with low molar mass polymers having small incompatibility — placing them in the weak segregation limit[136] — this effect can reasonably be considered negligible. Also, since mobility is essential for phase separation to occur, hydrophobic polymers with relatively low glass transition temperatures ($T_g$) are likely required. Polymersome morphology may also be influenced by other factors such as curvature, membrane thickness, copolymer architecture and the method applied for self-assembling polymers into vesicles. Specifically, the two-dimensional confinement of hydrophobic polymer chains within the vesicle membrane may introduce specific entropic constraints that are not accounted for in our simple Flory-Huggins approach. Nevertheless, due to the relatively large size of the polymersomes obtained in this study and the low molar mass of the polymers used, this effect appears to be negligible. When considering the potential scale-up of the process, some limitations may arise due to the electroformation method currently used. However, preliminary experiments using a double emulsion method have demonstrated that this alternative approach, which enables high yields, could also be employed effectively.

### CONCLUSION

Obtaining Janus vesicles presents a unique challenge: identifying a copolymer combination that is immiscible yet allows vesicle formation with both copolymers in the membrane. Our work demonstrates the significance and benefit of predicting and comparing system miscibility. By focusing on promising candidates, this approach facilitates the targeted exploration of relevant parameter space, leading to more efficient approach of Janus vesicle-forming systems. Our study confirms the relevance and reliability of the Flory-Huggins theory, originally developed for polymer miscibility in bulk, in predicting phase separation within vesicle membranes. We have achieved tunable phase separation by adjusting various parameters such as polymer mixture composition, degree of polymerization and temperature. Using this strategy, we have successfully conducted the self-assembly of two different block copolymers into asymmetric JGUVs, with both a high yield (more than 90 % of Janus-like vesicles) and a high stability over time. Overall, this work presents a significant contribution to the field of Janus vesicle design, opening avenues in areas such as nanorobots, microswimmers, active transport systems, drug delivery, and the study of active collective motion.

### MATERIALS AND METHODS

#### Materials

Poly(ethylene glycol)-*block*-poly(trimethylene carbonate) (PEG$_{22}$-*b*-PTMC$_{51}$), Mn= 6200 g.mol$^{-1}$ is synthetized according to a previously reported method.[135] Poly(ethylene glycol)-*block*-poly(trimethylene carbonate)-cyanine5.5 (PEG$_{22}$-b-PTMC$_{51}$-Cy5.5) is obtained from functionalization of PEG$_{22}$-b-PTMC$_{51}$-COOH with amino Cyanine 5.5. Poly(butadiene)-*block*-poly(ethylene glycol) (PBD$_{22}$–*b*-PEG$_{14}$), (P41745-BdEO, >85 % 1,2-addition of butadiene) and Poly(butadiene)-*block*-poly(ethylene glycol)-Rhodamine B (PBD$_{22}$–*b*-PEG$_{14}$-RhB), Mn=1800 g.mol$^{-1}$ (P9089A-BdEO-Rhodamine B, Mw/Mn 1.17, 89 % 1,2-addition of butadiene) have been purchased from Polymer Source, Inc. (Montreal, Canada). Poly(dimethylsiloxane)-*block*-poly(ethylene glycol) (PDMS$_{23}$-*b*-PEG$_{13}$, PDMS$_{27}$-*b*-PEG$_{17}$, PDMS$_{36}$-*b*-PEG$_{23}$) and poly(dimethylsiloxane)-nitrobenzoxadiazole (PDMS-NBD) have been synthesized and fully characterized in a previous work.[137] Characteristics of the copolymers used are provided in Table 1, Chlorofom (anhydrous, 99 %), sucrose and glucose have been purchased from Sigma-Aldrich.

#### GUVs preparation

Polymersomes are prepared using electroformation method reported by Angelova and Dimitrov.[138] Mixtures at different weight percent (w%) of two different block copolymers, each one with two w% of their corresponding labelled polymers with dyes, are prepared in chloroform at a total concentration of 1 mg.ml$^{-1}$. This solution is spread on the conductive faces of two indium tin oxide (ITO) coated glass slides. The slides are sealed on both sides of a rubber spacer using grease to form a chamber and dried under vacuum for 30 min. The ITO slides are connected to an AC voltage generator, a sinusoidal tension (2 V, 10 Hz) is applied and the chamber is filled with a 100 mM sucrose solution. The vesicles are collected after 1 hour (Figure 1). For solutions containing PTMC$_{51}$-*b*-PEG$_{22}$, the process is performed at 60°C, on a heating plate, to ensure a temperature above the melting point of the copolymer (~40°C).[135] For mixtures of PBD$_{22}$-*b*-PEG$_{14}$ with PDMS$_{27}$-*b*-PEG$_{17}$, electroformation is either made at room temperature or at 60°C. While, for clarity, composition is given in volume fraction (v%) of hydrophobic blocks in $\Delta G_m/n$ graphs, formulations are conveniently prepared based on w% of copolymers since these two values are closely related. The conversion between w% and v% is provided in Supporting Information Figure S1.

### Polymersomes extrusion

Polymersomes prepared by electroformation at 60°C with 50 w% of PBD$_{22}$-*b*-PEG$_{14}$ and 50 w% PDMS$_{27}$-*b*-PEG$_{17}$ are directly extruded with an Avanti® Polar Lipids Mini-Extruder, also at 60°C. 21 passes through a 5 µm, polycarbonate membrane from Genizer, are made to get nearly monodisperse vesicles.

### Confocal Observations

Vesicles solutions are redispersed into an equiosmotic glucose solution, in Ibidy® chambered coverslips for facilitating visualization of the vesicles at the bottom of the well due to density differences. All images are acquired using confocal laser scanning microscopy (CLSM, Leica TCS SP5, Leica Micro-systems CMS GmbH, Mannheim, Germany) inverted confocal microscope (DMI6000) equipped with a HCX PL APO x63 NA 1.4 oil immersion objective. PDMS-NBD (maximum excitation / emission: 458 nm / 523 nm), RhB-PEG-*b*-PBD (maximum excitation / emission: 546 nm / 567 nm) and Cy5.5-PTMC-*b*-PEG (maximum excitation / emission: 684 nm / 710 nm) are used as markers of each polymer and are sequentially imaged using three lasers: a 100 mW argon laser with an excitation at 488 nm and a range of emission at 495–530 nm; a 10 mW helium-neon laser with an excitation at 561 nm and a range of emission at 600–650 nm or 570–620 nm; and a 10 mW helium-neon laser with an excitation at 633 nm and a range of emission at 720–800 nm. Z-stacks of ~60 images covering a depth of ~30 µm at 1000 Hz were recorded for the 3D reconstruction of vesicles from electroformation. Concerning the vesicles extruded at 5 µm, Z-stacks of ~30 images covering a depth of ~25 µm at 8000 Hz in bidirectional mode have been achieved to reconstruct 3D images of the vesicles. The images are processed using ImageJ software.

### Statistical analysis

The percentage of vesicles being Janus like is determined by counting Janus and homogeneous vesicles on many z-stacks images using the multi-point tool on ImageJ software. Vesicle size measurements before and after extrusion is performed on ImageJ, measuring vesicle one by one. Measurements are made on z-projections of various z-stacks. Size distributions are fitted using a lognormal function (See Supplementary Information for details).

## ASSOCIATED CONTENT

**Supporting Information**.
Details on the calculation of Flory interaction parameter χ, correlation between w% of the copolymer and v% of the hydrophobic block and details on parameters and equations used for the polymersome size distribution characterization with supplementary tables S1 to S6 and supplementary figures S1 to S10. (PDF)
Supplementary Movies:
Movie S1: CLSM z-stack images of JGUVs (MP4)
Movie S2: CLSM z-stack reconstruction of JGUVs (MP4)
This material is available free of charge via the Internet at http://pubs.acs.org."

## AUTHOR INFORMATION

Corresponding Authors
**Eric Grelet** – CRPP, University of Bordeaux, CNRS, 33600 Pessac, France; Email: eric.grelet@crpp.cnrs.fr
**Sébastien Lecommandoux** – LCPO, University of Bordeaux, CNRS, Bordeaux INP, 33600 Pessac, France; Email: sebastien.lecommandoux@u-bordeaux.fr

Funding Sources

This work is supported by (a) MESRI, (b) Horizon Europe research and innovation program under the grant agreement No. 101079482 ("SUPRALIFE"); (c) ANR TEPEE (grant No. ANR-19-CE18-0024) and (d) ANR PopART (grant No. ANR-23-CE06-0008).

## ACKNOWLEDGMENT

The authors would like to thank Martin Fauquignon and Jean-Francois Le Meins for the synthesis of PDMS-b-PEG copolymers as well as Pierre Lalanne for the synthesis of PTMC-b-PEG copolymer. Anouk Martin is also acknowledged for fruitful discussions related to interaction parameters calculations. This work was funded by the European Union's Horizon Europe research and innovation programme under the grant agreement No. 101079482 ("SUPRALIFE"). The ANR TEPEE (grant No. ANR-19-CE18-0024) and ANR PopART (grant No. ANR-23-CE06-0008) are also acknowledged for financial support together with Univ. Bordeaux (RRI "Frontiers of Life"), CNRS and Bordeaux INP.

## ABBREVIATIONS

CLSM, Confocal laser scanning microscopy; Cy5.5, Cyanine 5.5; $f_h$, hydrophilic fraction; GUVs, giant unilamellar vesicles (diameter: 1-200 µm); ITO, Indium tin oxide; JGHUVs, Janus Giant Hybrid Unilamellar Vesicles; JGUVs, Janus giant unilamellar vesicles; LCST, lower critical solution temperature; LUV, large unilamellar vesicles (diameter: 100-1000 nm); Mn, number average molecular weight; NBD, nitrobenzoxadiazole; PBD, poly(butadiene); PDMS, polydimethylsiloxane; PEG, polyethylene glycol; PTMC, poly(trimethylene carbonate); RhB, Rhodamine B; T, temperature; Tm, melting temperature; Tg, glass transition temperature; v%, volume fraction; w% weight fraction; Đ, dispersity; δ, solubility parameter; $\Delta G_m/n$, molar Gibbs free energy of mixing; $\Phi$, volume fraction; $\chi_{AB}$, Flory interaction parameter.